\documentclass[aps,prl,sd,amsmath,amssymb,preprint,]
{revtex4-1}
\usepackage{graphicx}
\usepackage{dcolumn}
\usepackage{ulem}
\usepackage{xcolor}
\begin{document}

\title{On the Reactivity of Hydrogen-Helium and Hydrogen-Nitrogen at High Pressures}

\author{Robin Turnbull}
\affiliation{School of Physics and Astronomy and Centre for Science at Extreme Conditions, University of Edinburgh, Edinburgh, EH9 3FD, UK}

\author{Mary-Ellen Donnelly}
\affiliation{Center for High Pressure Science \& Technology Advanced Research, Shanghai, 201203, P.R. China}

\author{Mengnan Wang}
\affiliation{Center for High Pressure Science \& Technology Advanced Research, Shanghai, 201203, P.R. China}

\author{Cheng Ji}
\affiliation{Center for High Pressure Science \& Technology Advanced Research, Shanghai, 201203, P.R. China}
\affiliation{High Pressure Collaborative Access Team, Geophysical Laboratory, Carnegie Institution of Washington, Argonne, IL 60439}

\author{Phillip Dalladay-Simpson}
\affiliation{Center for High Pressure Science \& Technology Advanced Research, Shanghai, 201203, P.R. China}

\author{Miriam Pe{\~n}a-Alvarez}
\affiliation{School of Physics and Astronomy and Centre for Science at Extreme Conditions, University of Edinburgh, Edinburgh, EH9 3FD, UK}

\author{Ho-kwang Mao}
\affiliation{Center for High Pressure Science \& Technology Advanced Research, Shanghai, 201203, P.R. China}

\author{Eugene Gregoryanz}
\affiliation{School of Physics and Astronomy and Centre for Science at Extreme Conditions, University of Edinburgh, Edinburgh, EH9 3FD, UK}
\affiliation{Center for High Pressure Science \& Technology Advanced Research, Shanghai, 201203, P.R. China}

\author{Ross T. Howie*}
\affiliation{Center for High Pressure Science \& Technology Advanced Research, Shanghai, 201203, P.R. China}

\date{\today}

\begin{abstract}

Through a series of Raman spectroscopy studies, we investigate the behaviour of hydrogen-helium and hydrogen-nitrogen mixtures at high pressure across wide ranging concentrations. We find that there is no evidence of chemical association, miscibility, nor any demixing of hydrogen and helium in the solid state up to pressures of 250 GPa at 300 K. In contrast, we observe the formation of concentration-dependent N$_2$-H$_2$ van der Waals solids, which react to form N-H bonded compounds above 50 GPa. Through this combined study, we can demonstrate that the recently claimed chemical association of H$_2$-He can be attributed to significant N$_2$ contamination and subsequent formation of N$_2$-H$_2$ compounds. 

\end{abstract}

\maketitle

Understanding the behaviour of molecular mixtures under pressure is of a great importance to many scientific fields varying from chemistry to the studies of internal structures of astronomical bodies \cite{dewaele2016, loveday2001}. A wide range of phenomena have been observed in high-pressure molecular-mixtures such as 
phase separation, co-crystallisation, host-guest structures and chemical reaction \cite{morales2013, howie2016, degtyareva2007, guigue2017}. Since the discovery of solid van der Waals compounds in the high-pressure helium--nitrogen system \cite{vos1992}, binary mixtures of elemental gasses have attracted much attention. 	Subsequently, each binary mixture of the four lightest elemental gasses: H$_{2}$, He, N$_{2}$ and O$_{2}$ have been studied at high pressure \cite{ Loubeyre1985, Loubeyre1987, Loubeyre1991, Loubeyre1992, Loubeyre1995, Sihachakr2004, weck2010}. Recently, there has been renewed interest in studies of both the hydrogen-helium and hydrogen-nitrogen systems at high pressure investigating the synthesis of compounds through the reaction of the constituent molecules \cite{Lim2018, Spaulding2014, Wang2015, Goncharov2015, Laniel2018}. 

H$_2$ and helium are predicted to be chemically inert with one another, across a wide \textit{P-T} and concentration regime \cite{Ross1991, Ballone1995, Militzer2005, Bonev2007, Lorenzen2011, Ceperley2012}. Theoretical simulations motivated by potential miscibility within the Jovian planets, find evidence that even at these extreme conditions, hydrogen and helium are still phase separated. Due to the theoretical predictions of no chemical reactivity between hydrogen and helium, there have been few experimental studies on mixtures. Early studies exploring the eutectic phase diagram of hydrogen-helium mixtures found that in the two-fluid state the hydrogen intramolecular vibrational mode is markedly redshifted in He-rich concentrations, and was explained semiquantitatively by a helium compressional effect. \cite{Loubeyre1987}  However in the solid state, the two species were shown to be completely immiscible up to 15 GPa. This observation of immiscibility was utilized to grow single crystals of H$_2$, and measure the equation of state up to 100 GPa without an observable reaction between the two. \cite{Loubeyre1995} A recent high pressure study exploring H$_2$-He interactions as a function of mixture concentration, claimed the unprecedented appearance of hydrogen-helium solids at pressures below 75 GPa \cite{Lim2018}. Through the appearance of a vibrational Raman band at an approximate frequency to that calculated for the H-He stretch in a linear H-He-F molecule, the authors claim the formation of H-He bonds. \cite{Lim2018, Wong2000} These results are surprising given that it is a \textit{P-T} regime already explored both experimentally and theoretically. \cite{Loubeyre1987, Loubeyre1995}

\begin{figure}\includegraphics[width=0.6\columnwidth]{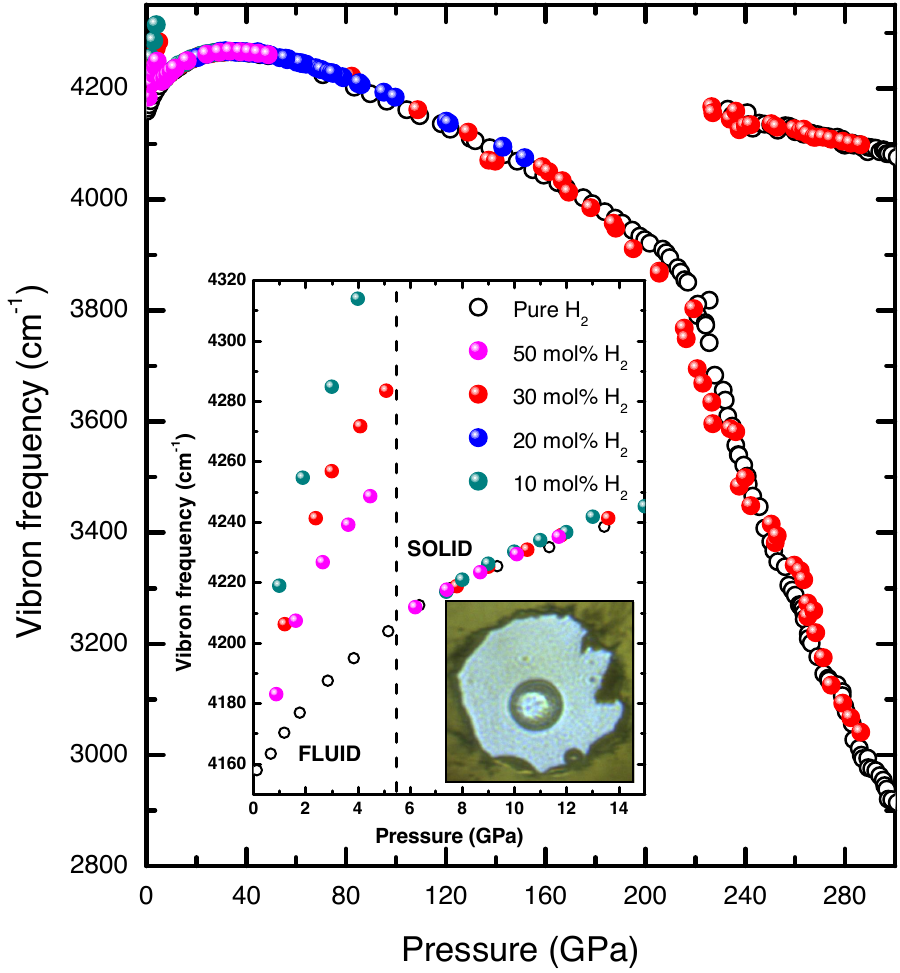}
\caption{(a) Raman frequency \textit{vs.} pressure for different H$_2$:He concentrations up to 285 GPa. Solid black line corresponds to pure H$_{2}$. Inset: Raman frequency \textit{vs.} pressure for different H$_2$:He concentrations up to 20 GPa. Photomicrograph of a sample mixture with 10 mol\% H$_2$ in He. The single crystal of H$_2$ is clearly phase separated from the He medium.}

\end{figure}

In contrast to H$_2$-He mixtures, the H$_2$-N$_2$ system exhibits particularly rich physics under compression, which is strongly dependant on both pressure and composition. Two van der Waals compounds have been reported to form at pressures above $\sim$ 7 GPa;  (N$_2$)$_6$(H$_2$)$_7$ and N$_2$(H$_2$)$_2$. Which compound, the pressure conditions at which the compounds are formed, and their characteristic Raman spectra are all dependent on the inital N$_2$:H$_2$ concentration \cite{Spaulding2014, Laniel2018}. At pressures between 35 - 50 GPa, these van der Waals compounds react to form amorphous solids. On decompression, the amorphous phase in all studies transforms to hydrazine (N$_{2}$H$_{4}$) on decompression below 10 GPa.\cite{Spaulding2014, Wang2015,Goncharov2015}

In this study, we have comprehensively investigated the reactivity of H$_2$-He and H$_2$-N$_2$ mixtures at high pressure as a function of mixture composition through Raman spectroscopy. Hydrogen and helium remain nearly immiscible across all concentrations up to pressures of 250 GPa, with no formation of van der Waals compounds nor any chemical reactivity across all mixture concentrations studied. Even at the extreme densities of hydrogen phase IV of hydrogen, which is thought to adopt a complex layered structure, no chemical association is observed. In contrast, modest pressures readily induces the formation H$_2$-N$_2$ van der Waals compounds, which with the application of higher pressure react to form ammonia, and on decompression, hydrazine. Through this combined study of both systems, we demonstrate that the recently reported chemical association between H$_2$ and He, can be described by the formation of N$_2$-H$_2$ compounds due to significant N$_2$ contamination of the H$_2$-He mixtures used in that study \cite{Lim2018}.

\begin{figure}\includegraphics[width=0.6\columnwidth]{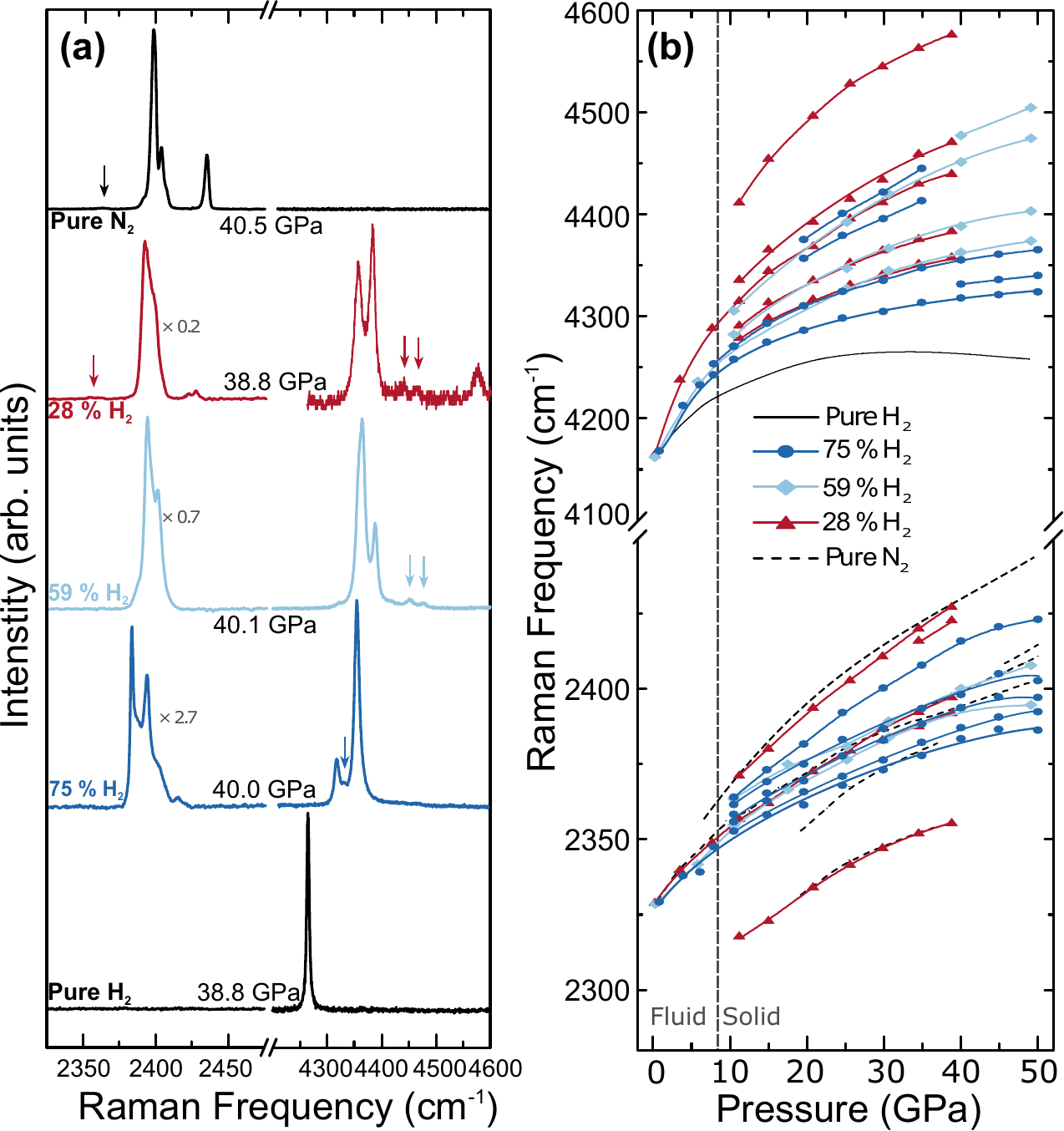}
\caption{(a) Evolution of the Raman spectra with composition the solid phase at around 40 GPa for the nitrogen-hydrogen compositions of:
	28, 
	59 and 
	75 mol\% H$_{2}$.
	The pure species are included for comparison in black.
	\label{fig:waterfall}
	(b) Raman frequency \textit{vs.} pressure plots corresponding to the data shown in (a).
	Solid and dashed black lines correspond to pure H$_{2}$ and pure N$_{2}$ respectively. Arrows indicate weak peaks.
	}
\end{figure}

Research grade (99.999\%) hydrogen-helium mixtures with molar hydrogen concentrations of 10, 20, 35 and 50\%. were obtained commercially (BOC).  Hydrogen--nitrogen compositions were prepared by ourselves from 99.999\% H$_2$ and N$_2$, with molar hydrogen contents of: 28, 50, 59 and 75 \% as determined from the relative partial pressures. The mixtures were given several days to homogenise before being gas loaded into diamond anvil cells (DACs).  All samples were gas loaded into the diamond anvil cells at a pressure of 200 MPa. Raman spectroscopy was conducted using 514 and 647 nm exitation wavelengths. Pressure was determined using the ruby fluorescence scale. \cite{dewaele2008}

Upon loading the samples of hydrogen-helium, all concentrations show only the Raman modes that can be atrributed to rotational modes (rotons) and vibrational modes (vibron) of H$_2$ (see Fig. S1). In the fluid, the two species are mixed well and the intensity of the hydrogen mode is constant when measured at different points across the sample chamber. However, the hydrogen vibron in the fluid is markedly red-shifted in frequency when compared with the pure species, and this red shift increases with increased helium concentration (see Fig.1). This is in good agreement with previous studies on the binary phase diagram. \cite{Loubeyre1987}

\begin{figure}
\label{fig1}
\includegraphics[width=0.6\columnwidth]{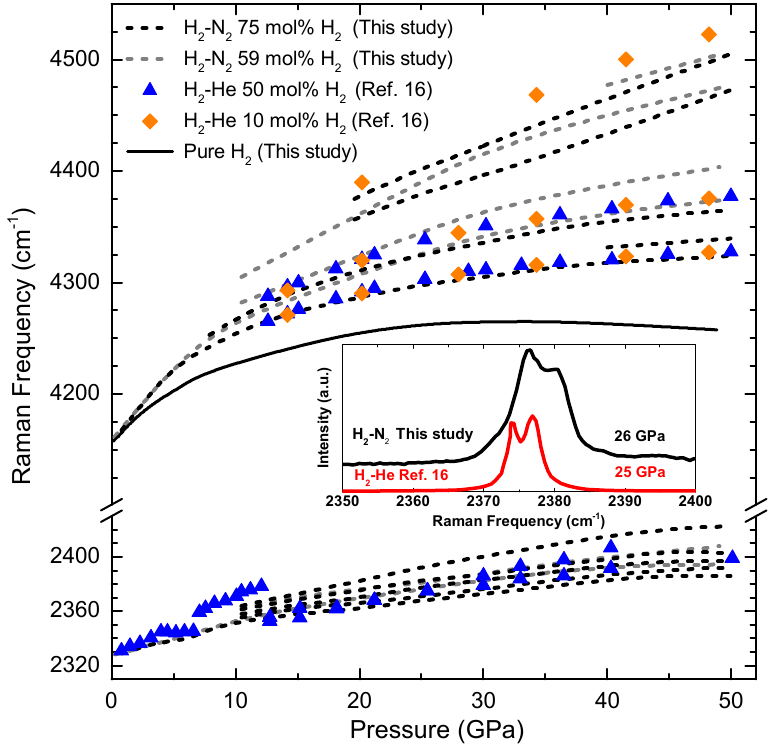}
\caption{Raman frequencies as a function of pressure for the claimed H$_2$-He S$_2$ compound of Ref. 1 (symbols) compared with our data on compounds formed in H$_2$-N$_2$ mixtures (dashed lines) and pure H$_2$ (black line). Inset: Comparison between `H-He' vibrons from Ref. 1 and the N$_2$ vibrons from a 3:2 H$_2$-N$_2$ mixture.}
\end{figure}

At pressures greater than 5.2 GPa, immiscibility becomes evident by the visible phase separation as hydrogen enters the solid state. With slow compression, the hydrogen crystals nucleate at the edge of the sample chamber and coallesce with time, whilst on rapid compression, we can form small H$_2$ crystallites across the whole sample chamber. In all concentrations studied, the Raman frequencies of the hydrogen vibron reverts to the same frequency as the pure species on crystallization. We still observe a weak H$_2$ vibron in the He media, indicating there are small crystallites of H$_2$ in the He medium, but there is neglible frequency difference compared with the bulk H$_2$. To rule out any kinetic effects, samples at each concentration were held in the fluid/solid states for a period of 1 week and no changes were observed with time. One sample at a concentration 20 mol\% H$_2$ was held for a period of 8 years at a pressure of 120 GPa with no evidence of a chemical reaction. 

At 300 K, pure hydrogen has been shown to go through a phase transition sequence of I-III at 180 GPa, and III-IV above 225 GPa. \cite{Howie2012} Phase IV is believed to adopt a two-layer molecular structure, giving rise to two distinct vibrational modes. One would expect that hydrogen in this phase would be more reactive, due to the much shorter molecular lifetime. It is also known that above 200 GPa, H$_2$ and D$_2$ tend to form molecular alloy with each other, unlike at lower pressures \cite{Howie2014} Fig. 1 (and Fig. S2) shows the vibron frequency as a function of pressure for a 30\% hydrogen in helium mixture up to the conditions of phase IV. We see only slight deviation in the vibron frequency when compared to pure H$_2$ and the deviation is well within experimental error of pressure determination. 

\begin{figure}[h]
\includegraphics[width=0.5\columnwidth]{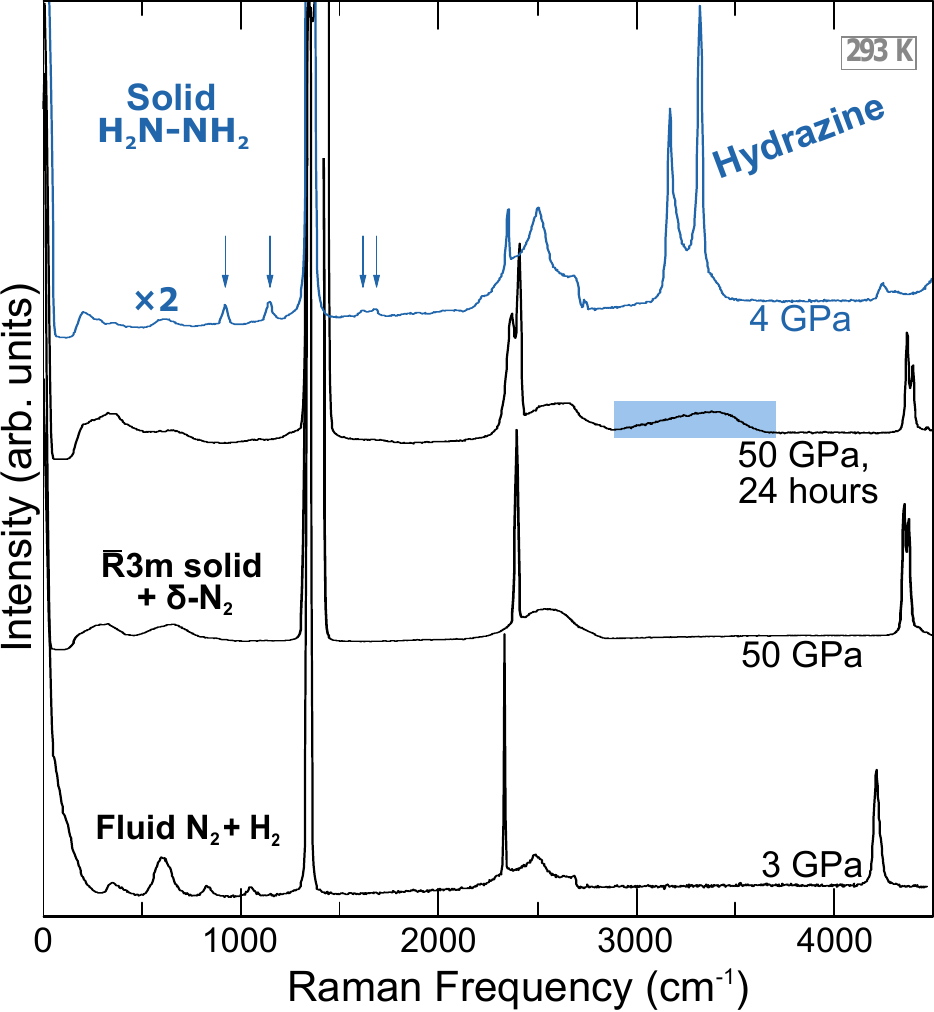}
\caption{Raman spectra on compression of a 50 mol\% H$_2$ composition demonstrating the time dependent chemical reaction of N$_2$ and H$_2$ above 50 GPa. Shaded blue region indicates the formation of N--H bonded compounds. Blue spectra shows the recovery of hydrazine on decompression to 4 GPa and the arrows highlight new, weak Raman peaks. \label{fig:azides}}
\end{figure}

Our data clearly shows that over a broad pressure regime, and over wide-ranging concentrations, there is no chemical interactions between H$_2$-He and they remain nearly imiscible up to pressures of 250 GPa. 
This is in direct contradicition of recent results reported in Ref. \cite{Lim2018} that claim chemical association between H$_2$-He in a He-rich fluid mixture. The evidence for this is primarily through the appearance of a Raman band at $\sim$2330 cm$^{-1}$ upon loading of the sample, which the authors attribute to an H-He bonded molecule. However, we do not observe this mode across all concentrations studied (see Fig. 1 and S1). N$_2$ has exactly the same vibrational frequency (2330 cm$^{-1}$ at 0.5 GPa) and it is unlikely that an H-He vibrational mode would have the same frequency dependency over a 10 GPa interval as a triple-bonded nitrogen molecule. Although the authors of Ref.  \cite{Lim2018} claim to rule N$_2$ contamination by the differences between N$_2$ (and N$_2$ in He) Raman vibrational frequencies in the solid state and their data, they do not consider the possiblities of N$_2$-H$_2$ interactions. As such, we present our own data, investigating the chemical interactions in H$_2$ and N$_2$ mixtures, in the pressure regime at which H$_2$-He chemical association is claimed. 

In the fluid state, hydrogen-nitrogen mixtures are characterized by two vibrational modes from H$_2$ and N$_2$ molecules (see Fig. S3). Across all concentrations studied, the vibron corressponding to H$_2$ molecules is red shifted and the shift, $\Delta\nu$, increases with greater N$_2$ concentration. In contrast, the N$_2$ vibron shows little effect by concentration in the fluid, and the pressure dependence follows closely with the pure species (see Fig. 2b). The solidification pressure of pure N$_2$ is 2 GPa, while it is 5.5 GPa for hydrogen. Interestingly, in mixtures of N$_2$ and H$_2$, all concentrations are homogeneous fluids below 8 GPa, before solidifying into van der Waals compounds. This is close to formation pressure claimed for the H$_2$-He compounds of Ref. \cite{Lim2018}. 

Fig.2a shows the Raman spectra of each concentration at $\sim$ 40 GPa compared with the pure species. At 75 mol\% H$_2$, we can identify the formation of both (N$_2$)$_6$(H$_2$)$_7$ and N$_2$(H$_2$)$_2$, through our powder x-ray diffraction measurements (see Fig. S4) and their characteristic spectra previously reported in Refs \cite{Spaulding2014, Laniel2018}. The coexistence of these compounds is different with respect to the previously reported binary phase diagram, which reported an overlap region between 54 mol\% and 66 mol\%. \cite{Spaulding2014} At lower H$_2$ concentrations of 59\% and 28\%, we see only (N$_2$)$_6$(H$_2$)$_7$ and $\delta$-N$_2$.

Fig. 3 compares the Raman frequencies of our 75 mol\% and 59 mol\% H$_2$-N$_2$ mixtures with that of 50 mol\% and 10 mol\% H$_2$-He mixtures in ref. \cite{Lim2018}. We find that the vibrational Raman modes of the claimed S$_2$ phase in a 5:5 H$_2$-He mixture matches closely to that of a 3:2 H$_2$-N$_2$ mixture across the whole pressure regime studied. At this mixture ratio, N$_2$(H$_2$)$_2$ is the dominant compound but co-exists with (N$_2$)$_6$(H$_2$)$_7$. At 12 GPa, the Raman mode at $\sim$4265 cm$^{-1}$ corresponds to the H$_2$ vibron in N$_2$(H$_2$)$_2$, whilst the higher frequency vibron at $\sim$4288 cm$^{-1}$ corresponds to the most intense H$_2$ vibrons in (N$_2$)$_6$(H$_2$)$_7$. In the high He content mixtures, which as a result includes higher N$_2$ content, there is a third H$_2$ vibrational mode at frequencies $\sim$4390 cm$^{-1}$ which corresponds the second most intense Raman band of (N$_2$)$_6$(H$_2$)$_7$. The behaviour of N$_2$ vibrons also behaviour very different in H$_2$-N$_2$ compounds than in either pure N$_2$ or N$_2$:He compounds. We find excellent agreement, shown in both Fig. 3 and the inset, between the N$_2$ stretches in N$_2$(H$_2$)$_2$ and the claimed `H$_2$-He' vibrational mode of Ref. \cite{Lim2018}. 
 
Above the critical pressure of 50 GPa at room temperature all samples exhibited loss of intensity of the hydrogen and nitrogen Raman vibrational bands over hour-long time-scales (see Fig. 4). This effect was also observed in Ref. \cite{Lim2018}, but interpreted as the formation of another H$_2$-He solid. The loss of vibron intensity occurred simultaneously with the emergence of a broad asymmetric peak centred around 3400 cm$^{-1}$ (highlighted in blue in Fig. 4). The broad asymmetric peak around 3400 cm$^{-1}$ can be attributed to the formation of N--H bonded vibrational modes. On decompression to below 10 GPa the broad peak around 3400 cm$^{-1}$ evolved into two sharp peaks accompanied by four lower frequency modes (see Fig. \ref{fig:azides}) unambiguously identifying hydrazine \cite{Jiang2014} Solid hydrazine was observed on decompression below 10 GPa in all isothermal compression-decompression experiments at room temperature up to 50 GPa. At pressures above 52 GPa, Ref. \cite{Lim2018} claim the formation of another H$_2$-He solid by the disappearance of the ``H-He'' and H$_2$ Raman bands, which is further evidence of N$_2$ contamination and the formation of N-H bonded compounds. 

The above analysis clearly shows that the recently claimed chemical association between H$_2$-He can be attributed to significant nitrogen contamination of the samples. The authors of Ref. 16 produce the gas mixtures themselves and it is most likely at the initial gas mixing stage in which the contaminant N$_2$ is introduced and the contamination increases with He concentration. We have extensive experience in producing gas mixtures and great care needs to be taken to ensure that the ballast volume between gas bottles in the mixture setup is adequately purged with the consituent gases.\cite{Howie2014, Howie2016} We have sometimes observed trace nitrogen contamination from air due to this, however in these cases we would disregard the contaminated gas bottle. In this study, we obtain our mixtures commercially with guaranteed levels of purity. Our results show, that even at extreme compressions, H$_2$ and He remain immiscible, a property which will prove advantageous for future structural studies of phase IV hydrogen. In agreement with previous theoretical results, it is likely that extreme \textit{P-T} conditions in excess of that in the interiors of Jovian planets would be required for H$_2$-He to become miscible, let alone form chemical bonds. 

\begin{acknowledgements}

Parts of this research were carried out at P02.2 at DESY, a member of the Helmholtz Association (HGF). We would like to thank H.-P. Liermann and K. Glazyrin for their assistance. M.P.-A. would like to acknowledge the support of the European Research Council (ERC) Grant “Hecate” Reference No. 695527. We would also like to thank M. Frost, V. Afonina for help in experiments and J. Binns for useful discussions. 

\end{acknowledgements}

\title{Supplementary Material for On the Reactivity of Hydrogen-Helium and Hydrogen-Nitrogen at High Pressures}

\author{Robin Turnbull}
\affiliation{School of Physics and Astronomy and Centre for Science at Extreme Conditions, University of Edinburgh, Edinburgh, EH9 3FD, UK}

\author{Mary-Ellen Donnelly}
\affiliation{Center for High Pressure Science \& Technology Advanced Research, Shanghai, 201203, P.R. China}

\author{Mengnan Wang}
\affiliation{Center for High Pressure Science \& Technology Advanced Research, Shanghai, 201203, P.R. China}

\author{Cheng Ji}
\affiliation{Center for High Pressure Science \& Technology Advanced Research, Shanghai, 201203, P.R. China}
\affiliation{High Pressure Collaborative Access Team, Geophysical Laboratory, Carnegie Institution of Washington, Argonne, IL 60439}

\author{Phillip Dalladay-Simpson}
\affiliation{Center for High Pressure Science \& Technology Advanced Research, Shanghai, 201203, P.R. China}

\author{Miriam Pe{\~n}a-Alvarez}
\affiliation{School of Physics and Astronomy and Centre for Science at Extreme Conditions, University of Edinburgh, Edinburgh, EH9 3FD, UK}

\author{Ho-kwang Mao}
\affiliation{Center for High Pressure Science \& Technology Advanced Research, Shanghai, 201203, P.R. China}

\author{Eugene Gregoryanz}
\affiliation{School of Physics and Astronomy and Centre for Science at Extreme Conditions, University of Edinburgh, Edinburgh, EH9 3FD, UK}
\affiliation{Center for High Pressure Science \& Technology Advanced Research, Shanghai, 201203, P.R. China}

\author{Ross T. Howie*}
\affiliation{Center for High Pressure Science \& Technology Advanced Research, Shanghai, 201203, P.R. China}

\date{\today}

\maketitle

\newcommand{\beginsupplement}{%
        \setcounter{table}{0}
        \renewcommand{\thetable}{S\arabic{table}}%
        \setcounter{figure}{0}
        \renewcommand{\thefigure}{S\arabic{figure}}%
     }
     \beginsupplement

\begin{figure*}
\includegraphics[width=0.5\columnwidth]{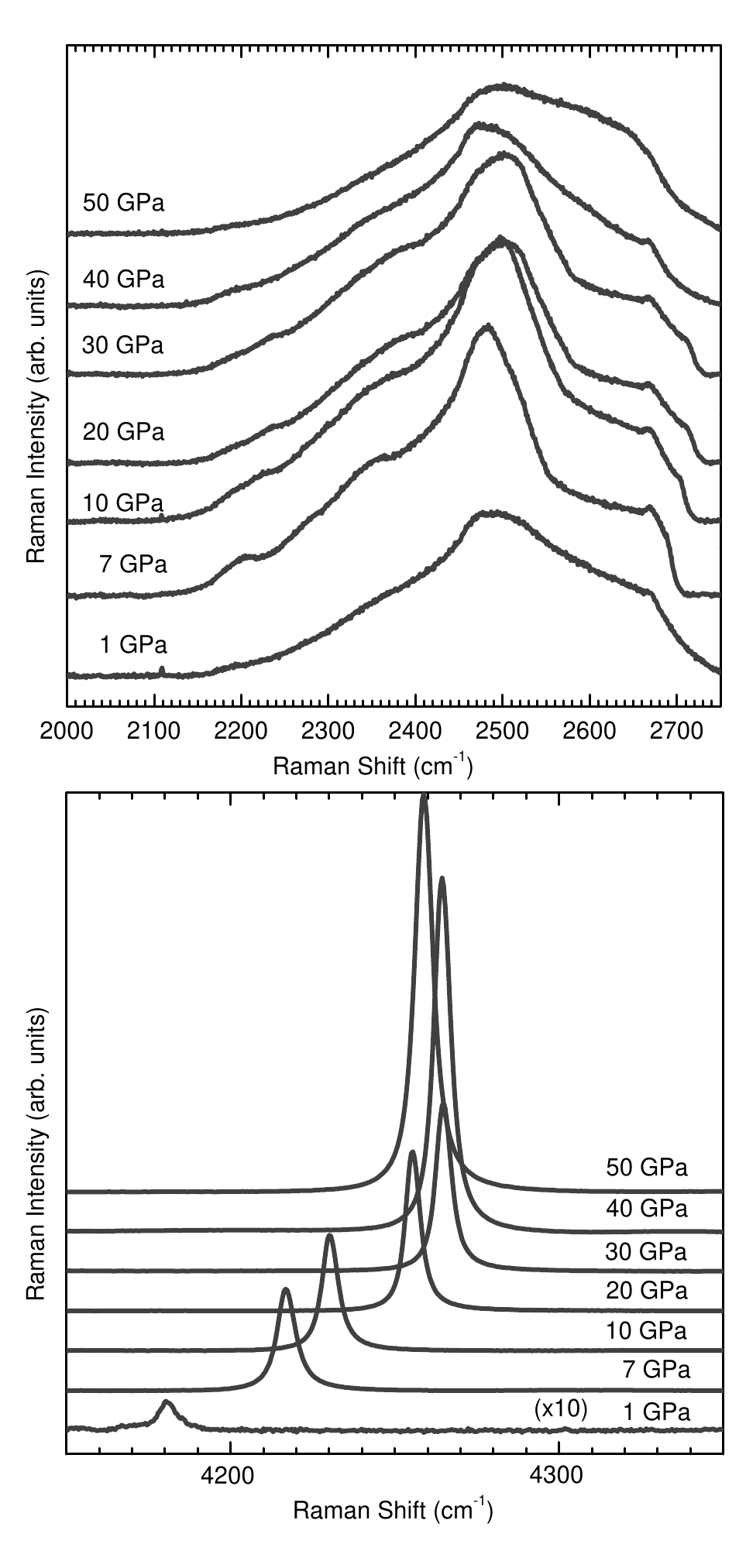}
\caption{Representative raw Raman spectra for a hydrogen-helium mixture of 10 mol\% H$_2$ as a function of pressure. The top panel shows the spectral region where a H$_2$-He mode was previously claimed in Ref.\cite{Lim2018}, which we do not observe in our study and attribute to N$_2$ contamination. The bottom panel shows the spectral region of the hydrogen intramolecular vibrational mode.}
\end{figure*}

\newpage

\begin{figure*}
\includegraphics[width=1\columnwidth]{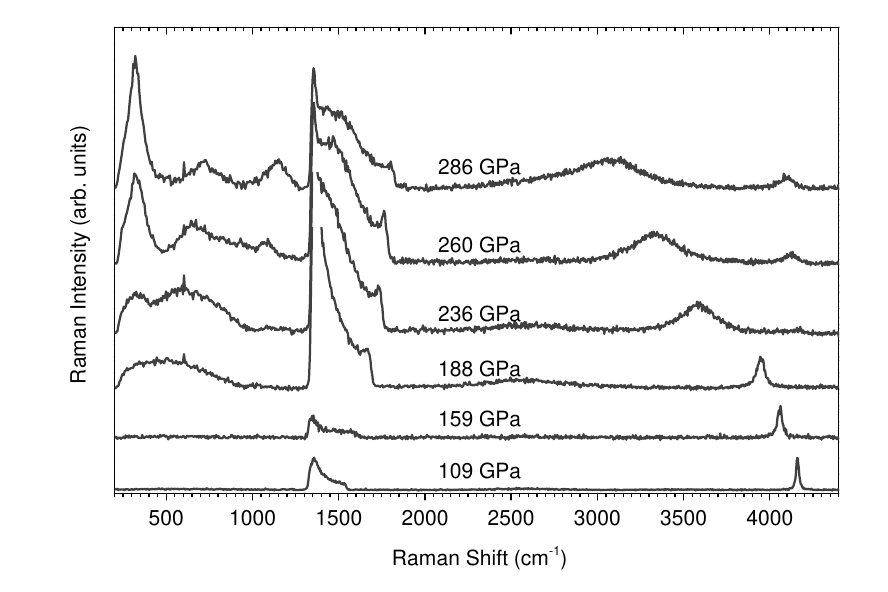}
\caption{Representative raw Raman spectra for a hydrogen-helium mixture of 20 mol\% H$_2$ as a function of pressure. The Raman spectra of the mixture has identical behaviour to that of the pure species. We observe the transition to H$_2$-IV above 220 GPa. No chemical association of H$_2$ and He is observed across the full pressure regime.}
\end{figure*}

\newpage

\begin{figure*}
\includegraphics[width=0.6\columnwidth]{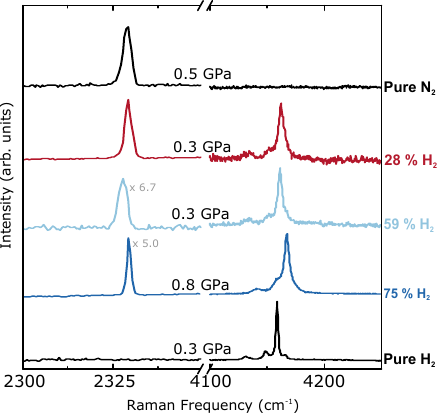}
\caption{Evolution of the Raman spectra with composition of H$_2$-N$_2$ mixtures in the fluid phase at approximately 0.5 GPa. The pure species are included for comparison in black. The individual spectra have been rescaled to a uniform intensity relative to the hydrogen vibron as indicated to allow easier comparison.}
\end{figure*}

\newpage

\begin{figure*}
\includegraphics[width=0.5\columnwidth]{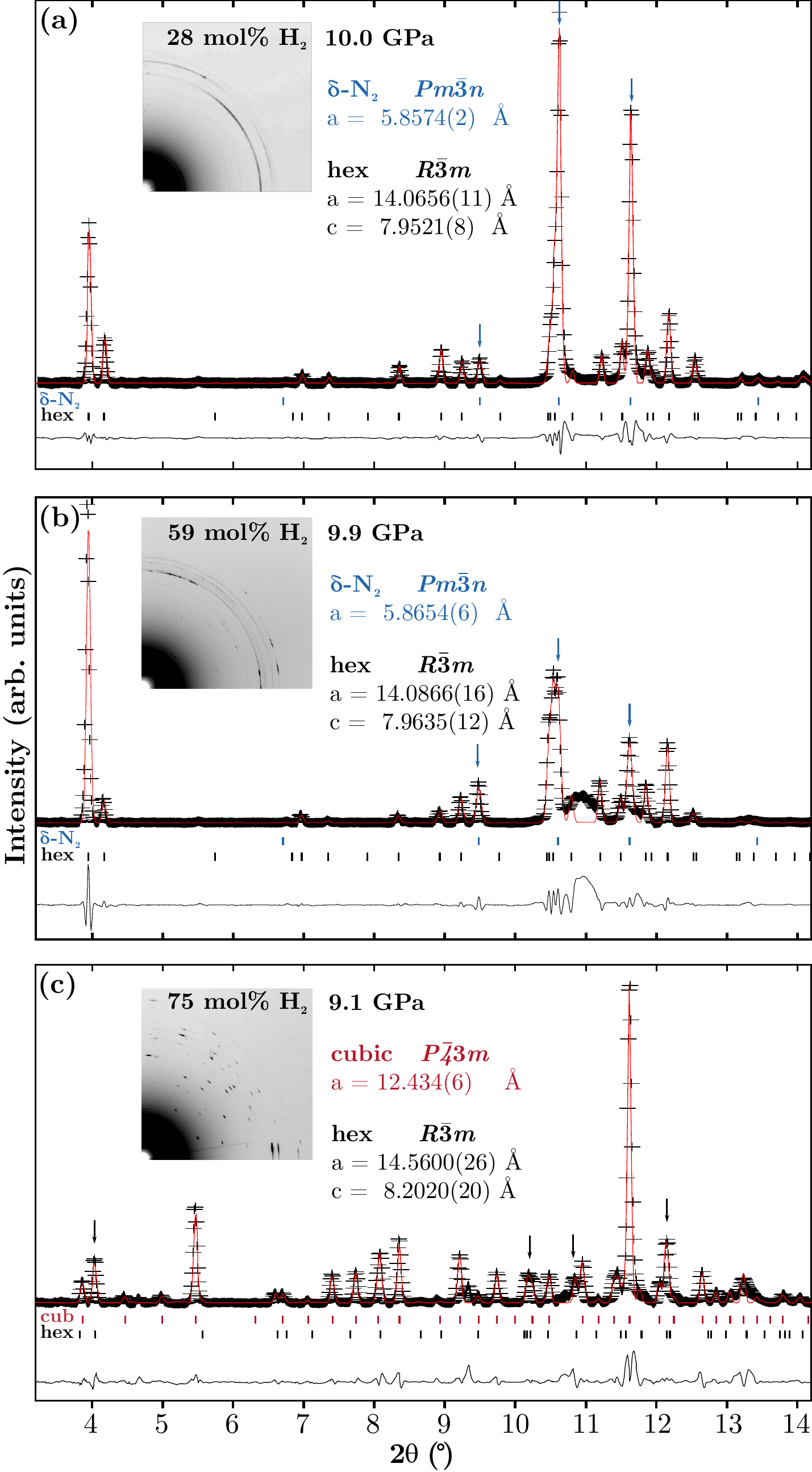}
\caption{Le Bail fits (red lines) to experimentally obtained x-ray powder diffraction data (black crosses) for three H$_2$-N$_2$ mixture samples of different concentrations:
	(a) 28 \% H$_{2}$,	
	(b) 58 \% H$_{2}$ and
	(c) 75 \% H$_{2}$.
	In (a) and (b) the peaks marked with blue arrows are assigned to $\delta$-N$_{2}$, the principal vibron of which is also detected in the Raman spectra. 
	All other peaks are well fitted by the \textit{R}$\bar{3}$\textit{m} hexagonal phase \cite{Spaulding2014}.
	In (c) the peaks marked with black arrows can only be fitted by the \textit{R}$\bar{3}$\textit{m} phase which has not been observed before in such a hydrogen-rich composition.  
	All others lines are indexed by a cubic \textit{P}$\bar{4}3$\textit{m} space group \cite{Laniel2018}.
	Insets: Raw diffraction image plates. The XRD was carried out at beamline P02.2 at PETRA III using a monochromatic beam of $\lambda$ = 0.484693 \r{A} focused to a spot size of $20 \times 20 \mu$m. 
	Data were recorded on a Mar 345 IP area detector and the Intensity vs. 2$\theta$ plots obtained by integrating the image plate data in DIOPTAS \cite{prescher2015}.}
\end{figure*}


\begin{references}

\bibitem{dewaele2016} A. Dewaele {\it et al.}, Nat. Chem., \textbf{8}, 784 (2016)

\bibitem{Morales2008} M.A. Morales {\it et al.}, PNAS, \textbf{106}, 1324 (2009)

\bibitem{loveday2001} J.S. Loveday {\it et al.}, Nature, \textbf{410}, 661 (2001)

\bibitem{morales2013} M.A. Morales {\it et al.}, Phys. Rev. B, \textbf{87}, 174105 (2013).

\bibitem{howie2016} R.T. Howie {\it et al.}, Sci. Rep. \textbf{6}, 34894 (2016).

\bibitem{degtyareva2007} O. Degtyareva {\it et al.}, Solid State Commun. \textbf{141}, 164 (2007).

\bibitem{guigue2017} B. Guigue {\it et al.}, Phys. Rev. B \textbf{95}, 020104 (2017).

\bibitem{vos1992} W. Vos {\it et al.}, Nature \textbf{358}, 46 (1992).

\bibitem{Loubeyre1985} P. Loubeyre {\it et al.}, Phys. Rev. B., {\bf 32}, 11 (1985).

\bibitem{Loubeyre1987} P. Loubeyre {\it et al.}, Phys. Rev. B., {\bf 36}, 7 (1987).

\bibitem{Loubeyre1991} P. Loubeyre, R. LeToullec, J. Pinceaux, J. Phys. Condens. Matt. \textbf{3}, 3183 (1991).

\bibitem{Loubeyre1992} P. Loubeyre {\it et al.}, Phys. Rev. B \textbf{45}, 22 (1992).

\bibitem{Loubeyre1995} P. Loubeyre {\it et al.}, Nature \textbf{378}, 44 (1995).

\bibitem{Sihachakr2004} D. Sihachakr {\it et al.}, Phys. Rev. B \textbf{70}, 134105 (2004).

\bibitem{weck2010} G. Weck {\it et al.}, Phys. Rev. B \textbf{82}, 014112 (2010).

\bibitem{Lim2018} J. Lim and C-S. Yoo, Phys. Rev. Lett., {\bf 120}, 165301 (2018).

\bibitem{Spaulding2014} D.K. Spaulding {\it et al.}, Nat. Comms., {\bf 5}, 5739 (2014).

\bibitem{Wang2015} H. Wang {\it et al.}, Sci. Rep., {\bf 5}, 13239 (2015).

\bibitem{Goncharov2015} A.F. Goncharov {\it et al.}, J. Chem. Phys., {\bf 142}, 214308 (2013).

\bibitem{Laniel2018} D. Laniel {\it et al.}, Phys. Chem. Chem. Phys., {\bf 20}, 4050 (2018).

\bibitem{Ross1991} J. E. Klepeis {\it et al.}, Science {\bf 254}, 986 (1991).

\bibitem{Ballone1995} O. Pfaffenzeller, {\it et al.} Phys. Rev. Lett. {\bf 74}, 2599 (1995).

\bibitem{Militzer2005} B. Militzer, J. Low Temp. Phys. {\bf 139}, 739 (2005).

\bibitem{Bonev2007} J. Vorberger {\it et al.}, Phys. Rev. B {\bf 75}, 024206 (2007).

\bibitem{Lorenzen2011} W. Lorenzen {\it et al.}, Phys. Rev. B {\bf 84}, 235109 (2011).

\bibitem{Ceperley2012} J.M. McMahon {\it et al.}, Rev. Mod. Phys. {\bf 84}, 1607 (2012).

\bibitem{Loubeyre1996} P. Loubeyre {\it et al.}, Nature, {\bf 383}, 702 (1996).

\bibitem{Wong2000} W. Wong {\it et al.}, J. Am. Chem. Soc. \textbf{122}, 6289 (2000).

\bibitem{dewaele2008} A. Dewaele {\it et al.}, Phys. Rev. B \textbf{78}, 104102 (2008).

\bibitem{Akahama2010} Y. Akahama {\it et al.}, J. Phys. Conf. Ser. \textbf{215},
012195 (2010).

\bibitem{Jiang2014} S. Jiang {\it et al.}, J. Phys. Chem. C \textbf{118},
3236 (2014).

\bibitem{Howie2012} R.T. Howie {\it et al.}, Phys. Rev. Lett. {\bf 108} 125501 (2012).

\bibitem{Howie2014} R.T. Howie {\it et al.} Phys. Rev. Lett. {\bf 113} 175501 (2014).

\bibitem{Ojwang2012} J. G. Ojwang {\it et al.}, J. Chem. Phys. \textbf{137}, (2012).

\bibitem{Ninet2016} S. Ninet {\it et al.}, J. Chem. Phys. \textbf{128}, 154508, (2016).

\bibitem{lundeen1975} J.W. Lundeen {\it et al.}, J. Phys. Chem. \textbf{79}, 2957 (1975).

\bibitem{haber1922} F. Haber, Naturwissenschaften \textbf{10}, 1041 (1922).

\bibitem{Liu2013} Q.J. Liu {\it et al.}, Comput. Theor. Chem., {\bf 1014}, 37 (2013).

\bibitem{Yin2015} K. Yin {\it et al.}, J. Mater. Chem. A, {\bf 3}, 4188 (2013).

 \bibitem{Hu2011} A. Hu {\it et al.}, J. Phys. Condens. Matter, {\bf 23}, 22203 (2011).
 
 \bibitem{Qian2014} G.-R Qian {\it et al.}, Sci. Rep., {\bf 6}, 25947 (2014).
 
 \bibitem{Steele2017} B.A. Steele and I.I. Oleynik, J. Phys. Chem. A, {\bf 121}, 1808 (2017).
 
 \bibitem{Batyrev2017} I. G. Batyrev, J. Phys. Chem. A, {\bf 121}, 638 (2017).
 
 \bibitem{Howie2016} R.T. Howie {\it et al.}, Sci. Rep., {\bf 6}, 34896 (2016).

\end{references}

\newpage
\clearpage

\begin{references}



\bibitem{Lim2018} J. Lim and C-S. Yoo, Phys. Rev. Lett., {\bf 120}, 165301 (2018).

\bibitem{Spaulding2014} D.K. Spaulding {\it et al.}, Nat. Comms., {\bf 5}, 5739 (2014).

\bibitem{Laniel2018} D. Laniel {\it et al.}, Phys. Chem. Chem. Phys., {\bf 20}, 4050 (2018).

\bibitem{prescher2015} C. Prescher and V. B. Prakapenka, High. Pres. Res., {\bf 35}, 223 (2015).



\end{references}
\end{document}